\documentclass[a4paper,11pt]{article}

\usepackage{amssymb}
\usepackage{amsmath}
\usepackage{graphicx}
\usepackage{comment}

\begin{document}

\huge

\begin{center}
Comment on ``A note on generalized radial mesh generation for plasma electronic structure''
\end{center}

\vspace{1cm}

\large

\begin{center}
Jean-Christophe Pain\footnote{jean-christophe.pain@cea.fr}
\end{center}

\normalsize

\begin{center}
CEA, DAM, DIF, F-91297 Arpajon, France\\
\end{center}

\vspace{1cm}

\begin{abstract}
In a recent note [High Energy Density Phys. \textbf{7}, 161 (2011)], B.G. Wilson and V. Sonnad proposed a very useful closed form expression for the efficient generation of analytic log-linear radial meshes. The central point of the note is an implicit equation for the parameter $h$, involving Lambert's function $W[x]$. The authors mention that they are unaware of any direct proof of this equation (they obtained it by re-summing the Taylor expansion of $h[\alpha]$ using high-order coefficients obtained by analytic differentiation of the implicit definition using symbolic manipulation). In the present comment, we present a direct proof of that equation.
\end{abstract}

The log-linear radial mesh is defined by the implicit equation \cite{wilson11}:

\begin{equation}\label{eq0}
kh=\alpha\frac{r_k}{r_c}+\ln\left[\frac{r_k}{r_c}\right].
\end{equation}

Substracting Eq. (\ref{eq0}) for $k$=1 from Eq. (\ref{eq0}) for $k$=$n$ gives

\begin{equation}\label{eq1}
(n-1)h=\alpha\frac{r_n-r_1}{r_c}+\ln\left[\frac{r_n}{r_1}\right].
\end{equation}

Since one has

\begin{equation}\label{eqh0}
h_0\triangleq\frac{1}{n-1}\ln\left[\frac{r_n}{r_1}\right]
\end{equation}

and

\begin{equation}
r_c\triangleq\frac{\alpha r_1}{W[\alpha e^h]},
\end{equation}

Eq. (\ref{eq1}) becomes

\begin{equation}\label{eq2}
\frac{(n-1)(h-h_0)~r_1}{r_n-r_1}=W[\alpha e^h].
\end{equation}

Equation (\ref{eqh0}) can be re-written as

\begin{equation}
r_n=r_1~e^{(n-1)h_0},
\end{equation}

which enables one to write Eq. (\ref{eq2}) as

\begin{equation}\label{eq3}
\frac{(h-h_0)}{d}~e^{h_0}=W[\alpha e^h],
\end{equation}

where $W$ represents Lambert's function and

\begin{equation}
d\triangleq\frac{e^{nh_0}-e^{h_0}}{n-1}.
\end{equation}

Equation (\ref{eq3}) is equivalent to

\begin{equation}
\frac{(h-h_0)}{d}~e^{h_0}~e^{\frac{(h-h_0)}{d}~e^{h_0}}=\alpha e^h,
\end{equation}

\emph{i.e.}

\begin{equation}
\frac{(h_0-h)}{d}~e^{(h_0-h)(\frac{d-e^{h_0}}{d})}=-\alpha,
\end{equation}

which leads, by multiplying both sides by $s\triangleq d-e^{h_0}=\frac{e^{nh_0}-ne^{h_0}}{n-1}$, to:

\begin{equation}
(h_0-h)\frac{s}{d}~e^{(h_0-h)\frac{s}{d}}=-\alpha s,
\end{equation}

which is equivalent to

\begin{equation}
h=h_0-\frac{d}{s}~W[-\alpha s],
\end{equation}

which is the result of Ref. \cite{wilson11}.

\end{document}